\documentclass[aps,prl,groupedaddress,]{revtex4}
\usepackage{latexsym}
\usepackage[dvips]{graphicx}

\newcommand{\eq}{\begin{equation}}
\newcommand{\feq}{\end{equation}}
\newcommand{\eqn}{\begin{eqnarray}}
\newcommand{\feqn}{\end{eqnarray}}
\newcommand{\arr}{\begin{eqnarray*}}
\newcommand{\farr}{\end{eqnarray*}}
\newcommand{\beq}{\begin{equation}}
\newcommand{\eeq}{\end{equation}}
\newcommand{\bea}{\begin{eqnarray}}
\newcommand{\eea}{\end{eqnarray}}

\def\beq{\begin{equation}}
\def\eeq{\end{equation}}
\def\feq{\end{equation}}
\def\bea{\begin{eqnarray}}
\def\eea{\end{eqnarray}}
\def\bc{\begin{displaymath}}
\def\ec{\end{displaymath}}
\def\lb{\label}

\def\ep{\epsilon}

\def\lb{\label}

\begin{document}


\title{Entanglement Entropy of two-dimensional anti-de Sitter black holes}

\author{Mariano Cadoni}
\email{mariano.cadoni@ca.infn.it}
\affiliation{Dipartimento di Fisica,
Universit\`a di Cagliari, and INFN sezione di Cagliari, Cittadella
Universitaria 09042 Monserrato, ITALY}


\begin{abstract}
Using the AdS/CFT correspondence we derive  a formula for the entanglement entropy 
of the anti-de Sitter black hole 
in two spacetime dimensions.  The  leading term in the large black 
hole mass expansion of our formula reproduces exactly the 
Bekenstein-Hawking  entropy $S_{BH}$, whereas  the 
subleading term behaves as $\ln S_{BH}$. 
This subleading term has the universal form 
typical for the entanglement entropy of physical systems described by 
 effective  conformal fields theories (e.g. one-dimensional 
 statistical models at  the critical point).
The well-known form of the 
entanglement entropy for a two-dimensional conformal field theory is 
obtained as analytic continuation of our result
and is related with  the entanglement entropy of a black hole with 
negative mass.

\end{abstract}


\maketitle

Quantum entanglement  is a fundamental feature of quantum systems.  
It is related to the existence of correlations between parts of the system. 
The degree of entanglement of a 
quantum system is measured by the entanglement  entropy $S_{ent}$.
In quantum field theory (QFT), or more in general in many body systems, we 
can localize observable and unobservable degrees of freedom in 
spatially separated regions $Q$ and $R$.
$S_{ent}$ is then  defined as the von Neumann entropy of the system when the degrees of 
freedom in the region $R$ are traced over, $S_{ent}=- 
Tr_{Q}{\hat \rho}_{Q}\ln\hat\rho_{Q}$, 
where the trace is taken over states in the observable region $Q$ and 
the reduced density matrix $\hat\rho_{Q}=Tr_{R}\hat\rho$ is obtained by tracing 
the density  matrix $\hat \rho$ over  states in the region $R$.

Investigation of the entanglement entropy (EE) has become relevant in many 
research areas. Apart from quantum information theory,  the field 
that gave birth to the notion of entanglement entropy, it plays a 
crucial role in condensed matter systems, where it helps to understand quantum 
phases of matter (e.g spin chains and quantum 
liquids)\cite{Vidal:2002rm,its,Kitaev:2005dm,Latorre:2004pk,korepin}.
Entanglement (geometric) entropy  is also an useful concept for 
investigating general features of QFT, in particular  
two-dimensional
conformal field theory (CFT) and  the Anti-de Sitter/conformal 
field theory (AdS/CFT) correspondence 
\cite{Holzhey:1994we,Calabrese:2004eu,Casini:2004bw,Fursaev:2006ng,
Solodukhin:2006xv,Ryu:2006bv,Ryu:2006ef} . 
Last but not least entanglement may held 
the key for unraveling the mystery of black hole entropy 
\cite{'tHooft:1984re,Bombelli:1986rw,Frolov:1993ym,
Fiola:1994ir,Belgiorno:1995xc,Hawking:2000da,Maldacena:2001kr,
Brustein:2005vx,Emparan:2006ni,Valtancoli:2006wv}.

We will be mainly concerned with the entanglement 
entropy of two-dimensional (2D)  CFT  and its relationship with the 
entropy of 2D black holes. It is an old idea that black hole entropy 
may be explained in terms of the EE  of the quantum state of 
matter fields in the  black hole geometry \cite{'tHooft:1984re}. The main 
support to this conjecture comes from the fact that both the EE 
 of matter fields and the Bekenstein-Hawking (BH) entropy depend on 
the area of the boundary region. On the other hand any attempt to 
explain the BH entropy as originating from quantum entanglement has 
to solve  conceptual and technical difficulties. 

The usual statistical paradigm explains the BH entropy in terms of 
a microstate gas. This   is conceptually different from the EE 
 that measures the observer's lack of information about the quantum 
state of the system in a inaccessible region of spacetime.
Moreover, the EE depends both on the number of species $n_{s}$ of the 
matter fields, whose entanglement should reproduce the BH entropy, and 
on the value of the UV cutoff $\delta$ arising owing to the presence of a sharp 
boundary between the accessible and inaccessible regions of the 
spacetime. Conversely, the BH entropy is meant to be universal, hence 
independent from $n_{s}$ and $\delta$.
Some conceptual difficulties can be solved  using Sakharov's 
induced gravity approach 
\cite{Jacobson:1994iw,Frolov:1996aj,Frolov:1997up}, but the problem of the dependence 
on $n_{s}$ and $\delta$ still remains unsolved.

In this letter we will show that in the case of two-dimensional AdS 
black hole these difficulties can be completely solved. We will 
derive an expression for the black hole EE that in the large black 
hole mass limit   reproduces exactly the 
BH entropy. Moreover, we will show that the subleading term has the 
universal behavior typical for CFTs and in particular for critical phenomena. 
The reason of this success is related to the 
peculiarities  of 2D AdS gravity, namely the existence of an 
AdS/CFT correspondence and the fact that  2D Newton constant can be 
considered as 
wholly induced by  quantum fluctuations of the dual CFT.

Most of the progress in understanding the EE in QFT has been 
achieved in the case of 2D CFT. Conformal invariance  in two 
space-time dimension is a powerful tool that allows us to compute the 
EE  in closed form.
The entanglement entropy for the ground state of a 2D CFT originated 
from tracing over 
correlations  between spacelike separated points has been calculated 
by  Holzhey, Larsen and Wilckzek \cite{Holzhey:1994we}.
Introducing an infrared cutoff $\Lambda$
the spacelike coordinate  of our  2D universe 
will belong to ${\cal C}= [0,\Lambda[¥$. The subsystem where 
measurements are performed is $Q=[0,\Sigma[$, whereas the 
outside region where the degrees of freedom are traced over is   
$R=[\Sigma, \Lambda[$. Because of the contribution of 
localized excitations 
arbitrarily near to the boundary the entanglement entropy diverges. 
Introducing an ultraviolet cutoff $\delta$,  the regularized 
entanglement entropy turns out to be \cite{Holzhey:1994we}
\beq\lb{f5}
S_{ent}= \frac{c+\bar c}{6}\ln\left(\frac{\Lambda}{\delta 
\pi}\sin\frac{\pi \Sigma}{\Lambda}\right),
\feq
where $c$ and $\bar c$ are the central charges of the 2D CFT. 
The expression (\ref{f5}) emphasizes the characterizing features of 
the entanglement entropy, namely subadditivity and invariance under the 
transformation which 
exchanges  the inside and outside regions
\beq\lb{f6}
\Sigma\to \Lambda-\Sigma.
\feq
Moreover, $S_{ent}$ is not a monotonic function of $\Sigma$, but 
increases and reaches its maximum for $\Sigma=\Lambda/2$ and then 
decreases as $\Sigma$ increases further.  This behavior has an 
obvious explanation. When the subsystem begins to fill most of the 
universe there is lesser information to be lost and the entanglement 
entropy decreases.

Let us now consider  2D AdS black holes.
As classical solutions of a 2D gravity theory they are endowed with 
a non-constant scalar field, the dilaton $\Phi$. 
In the Schwarzschild gauge the 2D AdS  black hole solutions 
are \cite{Cadoni:1994uf},
\beq\lb{f2}
ds^{2}= -\left(\frac{r^{2}}{L^{2}}-a^{2}¥\right)dt^{2}+
\left(\frac{r^{2}}{L^{2}}-a^{2}¥\right)^{-1}¥dr^{2},\quad \Phi= 
\Phi_{0}\frac{r}{L},
\feq
where the length $L$ is related to   
cosmological constant of the AdS spacetime ($\lambda=1/L^{2}$),
$\Phi_{0}$ is the dimensionless 2D inverse Newton constant and  $a$ is an
integration constants related to the 
black hole mass $M$ and horizon radius $r_{h}$ by
\beq\lb{f3}
a=\frac{r_{h}}{L}=\sqrt{\frac{2ML}{\Phi_{0}}}.
\feq
The thermodynamical, Bekenstein-Hawking, entropy of the black hole is 
\cite{Cadoni:1994uf} 
\beq\lb{f3a}
S_{BH}¥= 2\pi\Phi_{0}¥ a= 2\pi \sqrt{2\Phi_{0}ML},
\feq
whereas the black hole temperature is $T=a/2\pi L$.
Setting  $a=0$ in Eq. (\ref{f2})  we have the  AdS black hole ground 
state ( in the following called AdS$_{0}$) with zero mass, temperature 
and entropy. The AdS black hole (\ref{f2}) can be considered as the 
thermalization of the AdS$_{0}$ solution at  temperature $a/2\pi L$ 
\cite{Cadoni:1994uf}.

It has been shown that the 2D black hole has a dual description in 
terms of a
 CFT with central charge 
 \cite{Cadoni:1998sg,Cadoni:1999ja,Cadoni:2000kr,Cadoni:2000fq}
\beq\lb{f4}
c= 12\Phi_{0}.
\feq
The dual CFT can have  both the form of a 2D \cite{Cadoni:2000kr,Cadoni:2000fq}
or a 1D
\cite{Cadoni:1998sg,Cadoni:1999ja} conformal field  theory.
This  AdS$_{2}$/CFT$_{2}$ ( or AdS$_{2}$/CFT$_{1}$)
correspondence has been used to 
give a microscopical meaning to the thermodynamical 
entropy of 2D AdS black holes. Eq. (\ref{f3a})  has been reproduced 
by counting states in the  dual CFT.

In Ref. \cite{Fiola:1994ir} (see also Refs. \cite{Susskind:1994sm,
Frolov:1996aj,Frolov:1997up}) 
it was observed that in two dimensions  black hole entropy 
can be ascribed to quantum entanglement if 2D Newton 
constant  is wholly induced by  quantum fluctuations of matter fields.
On the other hand the AdS$_{2}$/CFT$_{2}$ correspondence, and in 
particular Eq. (\ref{f4}),  tells us that the 2D Newton constant 
is 
induced by quantum fluctuations of the dual CFT. 
It follows that the black hole entropy (\ref{f3a}) should be explained  
as the entanglement of the vacuum 
of the 2D CFT 
of central charge given by Eq. (\ref{f4}) in the gravitational black 
hole background (\ref{f2}).

At first sight one is tempted to  use Eq. (\ref{f5}) to calculate  
the entanglement 
entropy of the vacuum of the  dual CFT. 
The exterior region  of the 2D black hole 
can be easily identified with the region  $Q$, whereas 
the black hole interior has to be identified with the $R$  
region where the degrees of freedom are traced over.
There are two obstacles  that prevents   direct application 
of Eq. (\ref{f5}). First,  Eq. (\ref{f5}) holds for a 2D flat 
spacetime, whereas we  are dealing with  a curved 2D background.
Second, 
the calculations leading to Eq. (\ref{f5}) are performed for 
spacelike slice $Q$, whereas 
in our case 
the coordinate singularities  at  $r=r_{h}$ (the horizon) and $r=\infty$
(the timelike asymptotic boundary of the AdS spacetime)
do not allow for a global 
notion of spacelike coordinate (a coordinate system covering the 
whole black hole spacetime in which 
the metric is non-singular and static). 
Owing to these geometrical features, in the black hole
case we   cannot give a direct meaning to {\sl both} the measures 
$\Sigma$ and $(\Lambda-\Sigma)$ of the subsystems  $Q,R$.
As a consequence  invariance under the transformation 
(\ref{f6}) is meaningless in the black hole case.

The second difficulty can be circumvented using  appropriate 
coordinate system and  regularization procedure,  the first  using 
instead of Eq. (\ref{f5}) the formula derived by Fiola et al. 
\cite{Fiola:1994ir}, which gives the EE  of the vacuum of matter fields in the case of 
a curved gravitational background. 

In the coordinate system used to define the vacuum of scalar fields in 
AdS$_{2}$,
the 2D black 
hole metric (\ref{f2}) is \cite{Cadoni:1994uf}
\beq\lb{f9}
ds^{2}= 
\frac{a^{2}}{\sinh^{2}(\frac{a\sigma}{L})}\left(-dt^{2}+d\sigma^{2}\right).
\feq
The coordinate system $(t,\sigma)$  covers only the black 
hole  
exterior.  The black hole  horizon corresponds to $\sigma= \infty$ where 
the conformal factor of the metric vanishes.
The asymptotic $r=\infty$ timelike conformal boundary of the AdS$_{2}$ 
spacetime is located at $\sigma=0$, where the conformal factor 
diverges.  

The entanglement entropy of the CFT vacuum in the curved background 
(\ref{f9}) can be calculated, using the formula of Ref. 
\cite{Fiola:1994ir}
as the half line entanglement entropy seen by an observer in the 
$0<\sigma<\infty$ region. From the CFT point of view 
the AdS black hole has to be considered as the AdS$_{0}$ vacuum seen 
by the observer using the black hole coordinates (\ref{f9}) \cite{Cadoni:1994uf}.
Moreover, this observer sees the the AdS$_{0}$ vacuum as filled with 
thermal radiation with {\sl negative} flux \cite{Cadoni:1994uf}. It follows that 
the black hole entanglement entropy is given by the formula of Ref. 
\cite{Fiola:1994ir}  with reversed sign, 
\beq\lb{f7}
S_{ent}^{(bh)}¥=  -\frac{c}{6}\left( \rho(\sigma=0)- 
\ln\frac{\delta}{\Lambda }\right),
\feq
where $\rho$ defines the conformal factor of the metric in the 
conformal gauge ($ds^{2}= \exp(2\rho)(-dt^{2}+d\sigma^{2})$), $c$ is the 
central charge  given by Eq. (\ref{f4}) and 
$\delta,\Lambda$ are respectively UV and IR cutoffs.
Notice that in Eq. (\ref{f7}) we have only contributions from only one sector
(e.g. right movers) of the 
CFT. In Ref. \cite{Cadoni:2000kr,Cadoni:2000fq} it has been shown that the 2D  AdS black hole 
is dual to an open string with appropriate boundary conditions.
These boundary conditions are such that only one sector of the 
CFT$_{2}$ is present. The same is obviously true for the 
AdS$_{2}$/CFT$_{1}$ realization of the correspondence \cite{Cadoni:1998sg,Cadoni:1999ja}.

\begin{figure}
  \includegraphics[width=300pt]{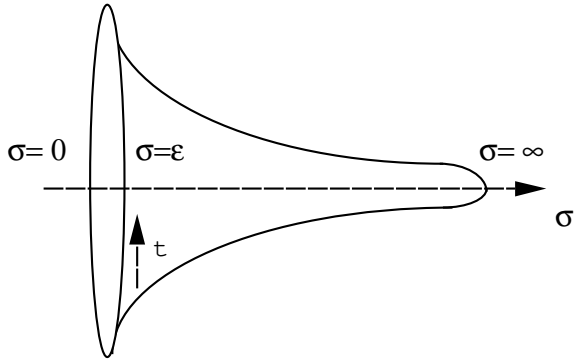}\\
  \caption{ Regularized euclidean instanton corresponding to the 2D 
  AdS black hole in the coordinate system $(t,\sigma)$ covering only 
  the black hole exterior. The euclidean time is periodic.
  The point $\sigma=\infty$ correspond to the 
  black hole horizon. $\sigma=0$ corresponds to the asymptotic 
  timelike boundary of AdS$_{2}$. }\label{instanton}
\end{figure}

The conformal factor of the metric (\ref{f9}), hence  the entanglement 
entropy (\ref{f7}) blows up on the 
$\sigma=0$ boundary of the AdS spacetime.    The simplest 
regularization procedure  that solves this problem is to consider a 
regularized boundary at $\sigma=\ep$.
Notice that $\ep$ 
plays the role of a UV cutoff for the coordinate $\sigma$, which is 
the natural spacelike coordinate of the dual CFT. $\ep$  is an IR 
cutoff for the coordinate $r$, which is the natural spacelike 
coordinate for the AdS$_{2}$ black hole.
The regularized euclidean instanton corresponding to the black hole 
(\ref{f9}) is shown in figure  (\ref{instanton}).
The regularizing parameter $\ep$ can be set equal to the  UV cutoff, 
$\delta=\ep$.
Moreover, the regularized boundary is at finite proper distance from 
the horizon so that $\ep$ acts also as IR regulator, making the 
presence of the IR cutoff $\Lambda$ in Eq. (\ref{f7}) redundant.
It follows that the regularized EE  is given by
$S_{ent}^{(bh)}¥=  -\frac{c}{6}\left( \rho(\ep)- 
\ln\frac{\ep}{L} \right)$, which using equations (\ref{f9}) and 
(\ref{f3}) becomes
\beq\lb{f11a}
S_{ent}^{(bh)}=\frac{c}{6} \ln \left( \frac{L^{2}¥}{r_{h}\ep} \sinh \frac{\ep 
r_{h}}{L^{2}}\right).
\feq
As a check of the validity of our formula we note that in the case of 
AdS$_0$ ($r_{h}=0$) the entanglement entropy vanishes.

The AdS/CFT correspondence enable us to identify the cutoff $\ep$
as the UV cutoff of the CFT : $\ep\propto L$.
The proportionality factor can be determined by requiring that the 
analytical continuation of Eq. (\ref{f11a})  is invariant under the 
transformation (\ref{f6})
(see later). This requirement fixes $\ep=\pi L$. With this position 
we get
\beq\lb{f11b}
S_{ent}^{(bh)}=\frac{c}{6} \ln \left( \frac{L¥}{\pi r_{h}} \sinh 
\frac{\pi 
r_{h}}{L}\right).
\feq
This formula is our  main result, it gives the entanglement entropy of 
the 2D AdS black hole.
This  entanglement entropy  has the expected  behavior as a 
function of the horizon radius $r_{h}$ or, equivalently, of the black 
hole mass $M$. $S_{ent}^{(bh)}$ becomes  zero in the AdS$_{0}$ ground 
state, $r_{h}¥=0$ ($M=0$), whereas it grows monotonically for $r_{h}>0$ 
($M>0$).

In order to compare  the black hole EE 
(\ref{f11b}) with the BH entropy (\ref{f3a}) let us 
consider the limit of macroscopic black holes, that is the limit 
$a\to \infty$ or equivalently $r_{h}>> L$ or also $M>>1/L$.
Expanding Eq. (\ref{f11b}) and using Eqs. (\ref{f3}) and (\ref{f4}) we 
get
\beq\lb{f12}
S_{ent}^{(bh)}=  2\pi 
\sqrt{2\Phi_{0}ML}-  \Phi_{0}¥\ln 
LM +O(1)= S_{BH}- 2\Phi_{0}\ln S_{BH} +O(1).
\feq
We have obtained the remarkable result that the leading term  in the 
large mass expansion of the black hole entanglement entropy reproduces exactly 
the Bekenstein-Hawking  entropy. Moreover, the subleading term 
behaves as the logarithm of the BH entropy and describes quantum 
corrections to $S_{BH}$.
It is an universally accepted result that the quantum corrections to 
the BH entropy behave as $\ln S_{BH}$ 
\cite{Fursaev:1994te,Mann:1997hm,Kaul:2000kf,Carlip:2000nv,Ghosh:1994wb,
Mukherji:2002de,Setare:2003vv,Domagala:2004jt,Medved:2004eh,Grumiller:2005vy}.
However, there is no 
general consensus about the value of the prefactor of this term. For the 
microcanonical ensemble this term has to be negative,  whereas there are 
positive contributions coming from thermal fluctuation. 
Equation (\ref{f12}) fixes the prefactor of  $\ln S_{BH}$ in terms 
of the  2D Newton constant. This result 
contradicts some previous results  supporting a 
$\Phi_{0}$-independent  value of the prefactor.
Our result is consistent with the  approach followed in this paper,
which considers 2D gravity as induced from the quantum fluctuations 
of a  CFT with central charge $12\Phi_{0}$. 
The first (Bekenstein-Hawking) term in Eq. (\ref{f12}) is the 
induced entanglement
entropy, whereas the second term, $-( c/6)\ln (r_{h}/L)$, is determined by the conformal 
symmetry.  It gives the entanglement 
entropy (\ref{f5})  of a CFT in 2D flat spacetime  with central charge
$12\Phi_{0}$ and 
$\Sigma=r_{h}$ in the limit $\Sigma <<\Lambda$ \cite{Holzhey:1994we}. 
The subleading term in Eq. (\ref{f12}) represents therefore an 
universal 
behavior shared  with other  systems described by 2D QFTs,
such as one-dimensional 
statistical models near  to the critical point (with  the black hole 
radius $r_{h}$
corresponding to the  correlation length) or free  scalars 
fields  
\cite{Calabrese:2004eu,Fursaev:2006ng}.

Eq. (\ref{f11b}) shows a close 
resemblance with the  CFT  entanglement entropy (\ref{f5}).
Eqs. (\ref{f11b}) and (\ref{f5}) differs in two main points: the 
absence in the black hole case of something corresponding to the 
measure of the whole space (the parameter $\Lambda$ in Eq. (\ref{f5})) 
and the appearance of  hyperbolic instead of 
trigonometric  functions. 
These  are expected features for the entanglement entropy 
of a black hole. They solve the  problems concerning 
the application of  formula (\ref {f5}) to the black hole case.
For a black hole one cannot define a measure of  the whole space 
analogue to $\Lambda$. For static solutions the coordinate system 
covers only the black hole exterior.
The  appearance of hyperbolic  instead of trigonometric functions 
allows for monotonic increasing of $S_{ent}^{(bh)}(r_{h})¥$, 
eliminating the  unphysical decreasing  
behavior of $S_{ent}(\Sigma)$ in the region $\Sigma> \Lambda/2$.

\begin{figure}
  \includegraphics[width=300pt]{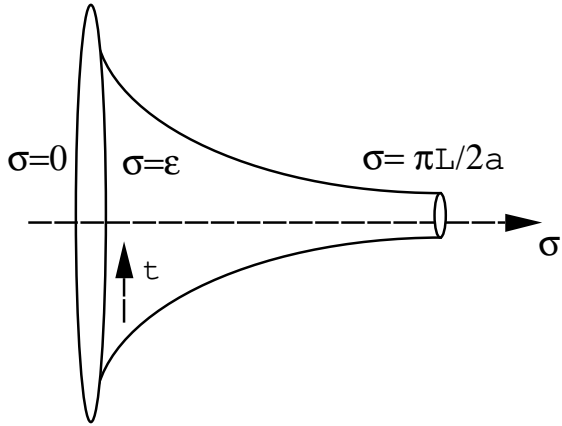}\\
  \caption{ Regularized euclidean instanton corresponding to the 2D 
  AdS black hole with negative mass. The euclidean time is periodic.
  The point $\sigma=\pi L/2a$ 
  corresponds to the 
  black hole singularity at $r=0$. $\sigma=0$ corresponds to the asymptotic 
  timelike boundary of AdS$_{2}$. }\label{instanton1}
\end{figure}

It is interesting to see how Eq. (\ref{f5}) can be obtained as the 
analytic continuation $r_{h}\to i r_{h}$ of our formula (\ref{f11b}), 
i.e  by considering an AdS black hole with negative mass.
The analytically continued black hole solution is given by Eq. 
(\ref{f2}) with $a^{2}<0$. In the conformal gauge the solution 
reads now $ds^{2}= [a^{2}/\sin^{2}(a\sigma /L)](-dt^{2}+d\sigma^{2})$.
The range of the spacelike coordinate, corresponding to $0< r<\infty$, 
is now  $0<\sigma< \pi L/2a$. Regularizing the solution at 
$\sigma=0$ by introducing the cutoff $\ep$ we get the euclidean 
instanton shown in Fig. (\ref{instanton1}). 
In terms of the 2D CFT we have to trace over 
the degrees of freedom outside the spacelike slice $\ep <\sigma< 
\pi L/2a$. The related entanglement entropy can be calculated using the 
formula of Ref. \cite{Fiola:1994ir} in the case of a spacelike slice  with 
two boundary points: $S_{ent}= -c/6 
[\rho(\ep)+\rho(\pi L/2a)-\ln(\delta/\Lambda)]$.  Applying this formula 
to the case of the black hole solution of negative mass, identifying
 $\ep$ 
 in terms of the IR cutoff $\Lambda$, $\ep= \pi L^{2}/\Lambda$, and 
 redefining appropriately the UV cutoff $\delta$, we get
\beq\lb{f15}
S_{ent}= \frac{c}{6} \ln \left(\frac{\Lambda}{\pi \delta} \sin\frac{\pi 
r_{h}}{\Lambda}\right).
\feq
Thus, the entanglement entropy of the 2D CFT in the curved background 
given by the AdS black hole of negative mass has exactly the form 
given by Eq. (\ref{f5}) with the horizon radius $r_{h}$ playing the 
role of $\Sigma$.
Notice that the presence of the factor $\pi$ in the argument of the 
 $\sin$-function  is necessary if one wants invariance under the transformation 
 (\ref{f6}). The requirement that  equation (\ref{f15}) is the analytic 
 continuation of Eq. (\ref{f11b})  fixes, as previously 
 anticipated, the proportionality factor between $\ep$ and $L$ in the 
 calculations leading to Eq. (\ref{f11b}).
 
 In this letter we have derived a formula for the entanglement 
 entropy of 2D  AdS black holes that has nice striking features.
 The leading term in the large black hole mass expansion reproduces 
 exactly the BH entropy. The subleading term has the right $\ln 
 S_{BH}$, behavior of the quantum corrections to the BH 
 formula and represents an universal term typical of CFTs.
 Analytic continuation to negative black hole masses 
 give exactly the entanglement entropy of 2D CFT with the black hole radius 
 playing the role of the measure of the observable spacelike slice in the 
 CFT. 
 Our results rely heavily on  peculiarities of 2D AdS gravity, namely 
 the  existence of an AdS/CFT correspondence and on the fact that 2D 
 Newton constant arises from quantum fluctuation 
 of the dual CFT.
 The generalization of our approach to higher dimensional gravity 
 theories is therefore far from being trivial. A related problem is the 
 form of the coefficient of the $\ln 
 S_{BH}$ term. In the 2D context our result, stating that this 
 coefficient is given in terms of the 2D Newton constant (or 
 equivalently the central charge of the  dual CFT) is rather 
 natural. For higher dimensional gravity theories this is again a 
 rather subtle point.

\acknowledgements
I thank G. D'Appollonio  for discussions and valuable comments.

\end{document}